\documentclass[prl,aps,twocolumn]{revtex4}
\usepackage{graphicx}
\usepackage{amssymb,amsmath}
\usepackage{bm}
\newcommand{\kk}{\bm{k}}
\newcommand{\qq}{\bm{q}}
\newcommand{\bk}{b_{\bm{k}}}

\newcommand{\bkd}{b_{\bm{k}}^{\dag}}
\newcommand{\bmkd}{b_{-\bm{k}}^{\dag}}
\newcommand{\bek}{\beta_{\bm{k}}}
\newcommand{\bekd}{\beta_{\bm{k}}^{\dag}}
\newcommand{\bemkd}{\beta_{-\bm{k}}^{\dag}}

\newcommand{\beqd}{\beta_{\bm{q}}^{\dag}}
\newcommand{\sumk}{\sum_{\bm{k}}}
\newcommand{\fk}{f_{\bm{k}}}
\newcommand{\Fk}{F_{\bm{k}}}
\newcommand{\gk}{g_{\bm{k}}}
\newcommand{\wk}{\omega_{\bm{k}}}
\newcommand{\wq}{\omega_{\bm{q}}}
\newcommand{\pe}{|1\rangle\!\langle 1|}
\newcommand{\pg}{|0\rangle\!\langle 0|}
\newcommand{\rl}{\rangle\!\langle}

\begin{document}

\author{Pawe{\l} Machnikowski}
\email{Pawel.Machnikowski@pwr.wroc.pl}
\affiliation{Institute of Physics, Wroc{\l}aw University of
Technology, 50-370 Wroc{\l}aw, Poland}

\title{Change of decoherence scenario and appearance of localization
due to reservoir anharmonicity}

\begin{abstract}
Although coupling to a super-Ohmic bosonic reservoir leads only to
partial dephasing on short time scales, exponential decay of coherence
appears in the Markovian limit (for long times) if anharmonicity of
the reservoir is taken into account. This effect not only
qualitatively changes the decoherence scenario but also leads to
localization processes in which superpositions of spatially separated 
states dephase with a rate that depends on the distance 
between the localized states. As an example of the latter process, we
study the decay of coherence of an electron state delocalized over two
semiconductor quantum dots due to anharmonicity of phonon modes.
\end{abstract}


\maketitle

Decoherence of open quantum systems has become one of the central
issues of the quantum theory. On one hand, erasure of
phase information with respect to a certain basis of states
may bring classical behavior out of quantum evolution
\cite{joos03,zurek93,zurek03}. 
On the
other hand, the ability to control the states of quantum systems may
open the way to novel applications, like quantum information processing
\cite{nielsen00,alber01}, provided that quantum coherence is maintained over a
sufficiently long time. Therefore, understanding
and reducing decoherence is also of practical importance.

Of special interest is a class of models that allow 
for erasure of
phase information without transitions between the selected
basis states. Such a \textit{pure dephasing} process is an
essential ingredient of a measurement, with the basis states
selected by the coupling between the measurement device and its
environment (the \textit{pointer basis}) \cite{zurek81} and
determining the physical meaning of the measurement \cite{zurek03}. 
A simple model leading to this kind of behavior is composed of a
two-level system and a bosonic bath with the coupling between these
two subsystems linear in the bosonic operators and commuting with the
Hamiltonian of the system (\textit{independent boson model}
\cite{mahan00}). 
Pure dephasing effects are also relevant
for the short time dynamics of confined carrier states in
semiconductor quantum dots (QDs) \cite{krummheuer02,vagov03}. 
A class of systems described by such a model 
(including confined carriers in QDs interacting
with phonons) shows only \textit{partial} dephasing with a
finite asymptotic level of coherence \cite{krummheuer02,alicki04b}.
This feature of real systems
is essential for the possibility of explaining 
the classical nature of measurement results in terms of dephasing 
(``einselection'')
\cite{zurek03}, which is based on the expectation that dephasing
processes are asymptotically exponential, as observed in some formal
models \cite{zurek82}.

The purpose of this paper is to show that the important property of
complete or partial asymptotic dephasing depends not only on the
system-reservoir coupling but also on the properties of the reservoir
itself. In particular, no real reservoir may be strictly harmonic, one
obvious reason being the need to restore the equilibrium, i.e., a
sufficient level of ergodicity. Here, this is accounted for by
including an anharmonic coupling between the bosonic modes. This model
closely corresponds to real properties of solid-state phonon
reservoirs, where also parameters of the model may be inferred from
experiments. 

The combination of system-reservoir
coupling with reservoir anharmonicity leads to a dephasing effect that,
for spatially localized systems, may be understood in terms of
collisional decoherence: 
Because of their coupling to the system, reservoir modes undergo
a shift of their equilibrium positions which depends on the system
state, forming a coherent displacement field \cite{jacak03b,vagov02a}.
If the modes are coupled by
anharmonicity the displacement field acts as a scattering
potential for other reservoir modes. Since the displacement is
state-dependent, each scattering event extracts a certain amount of
information on the system state, gradually leading to complete 
dephasing. This
is the first essential result of the present work.
The second result is related to the special situation when the
system is in a superposition of two states corresponding to distinct
positions in real space. 
In such case, scattering of reservoir modes leads to vanishing of the
coherence between the two distant states.
This process turns a genuine quantum-delocalized system state into a
mixture of two classical-like localized states and is hence referred
to as \textit{localization}  \cite{joos85,gallis90,hornberger03}.
It is shown, for a specific model of
carrier-phonon interaction in semiconductor QDs,
that the rate of localization grows
with the separation between the two spatial locations. 
This confirms
the intuitive expectation that coherence on large distances should
be more fragile.

Let us consider the Hamiltonian $H=H_{0}+H_{1}$, where $H_{0}$ is the
two-level independent-boson Hamiltonian
\begin{equation}\label{h0}
    H_{0}=-\frac{1}{2}E\sigma_{z}+\sumk \hbar \wk\bkd\bk
    +\frac{1}{2}\sigma_{z}\sumk \Fk^{*}(\bk+\bmkd),
\end{equation}
with $\Fk^{*}=F_{-\bm{k}}$, 
and $H_{1}$ describes the third order anharmonic coupling between
various phonon modes
\begin{equation}\label{h1}
H_{1}=\frac{1}{6}
\sum_{\kk_{1}\kk_{2}\kk_{3}} 
w_{\kk_{1},\kk_{2},\kk_{3}} \delta_{\kk_{1}+\kk_{2}+\kk_{3}=0}
A_{\kk_{1}}A_{\kk_{2}}A_{\kk_{3}},	
\end{equation}
where $A_{\kk}=\bk+\bmkd$ and the anharmonic constants 
$w_{\bm{k}_{1},\bm{k}_{2},\bm{k}_{3}}=
w^{*}_{-\bm{k}_{1},-\bm{k}_{2},-\bm{k}_{3}}$ are symmetric
under permutation of indices. The polarization (branch) 
index of the boson modes is implicit in $\bm{k}$.

We define the unitary operator
\begin{equation}\label{weyl}
\mathbb{W}=\pg\otimes W^{\dag} +\pe\otimes W,
\end{equation}
where 
$W=\exp[ (1/2)\sumk \gk^{*}\bk-\mathrm{H.c.}]$, 
$\gk=\Fk/(\hbar\wk)$.
In terms of the new operators
$\bek=\mathbb{W}\bk\mathbb{W}^{\dag}=\bk+\frac{1}{2}\sigma_{z}\gk$,
the Hamiltonian $H_{0}$ is diagonal, 
$H_{0}=-(1/2)E\sigma_{z}+\sumk \hbar\wk\bekd\bek$. Using the exact
diagonalization by the operator $\mathbb{W}$ one can find the
evolution of the non-diagonal element of the reduced density matrix
(in the original basis), 
$|\rho_{01}(t)|=|\rho_{01}(0)|\exp[
-2\!\!\int\!\! d\omega
\coth(\hbar\omega/k_{\mathrm{B}}T)
J(\omega)\sin^{2}(\omega t/2)/(\hbar\omega)^{2}]$,
where $J(\omega)=\sumk|\Fk|^{2}\delta(\omega-\wk)$. For
spectral densities sufficiently regular at low frequencies, 
$J(\omega)\sim\omega^{n}$, $n\ge 2$ (\textit{super-Ohmic} reservoirs)
and for gapped reservoirs ($J(\omega)=0$ around $\omega=0$),
$|\rho_{01}|$ reaches a finite asymptotic value
(corresponding to partial dephasing)
\cite{krummheuer02,alicki04b}. On the other hand, if 
$J(\omega)\sim\omega$
(\textit{Ohmic} reservoirs), $\rho_{01}$ decays exponentially for long times
\cite{breuer02}.
It turns out, however, that Ohmic independent boson models show
infrared divergences that cause problems on the formal level
\cite{alicki04b}. Below we will see that dephasing generated by a
super-Ohmic reservoir becomes complete (exponential at long times) if
the reservoir is anharmonic.

In terms of the transformed operators $\beta$ 
the anharmonic Hamiltonian $H_{1}$ becomes
\begin{eqnarray*}
H_{1} & = & \frac{1}{6}
\sum_{\kk_{1}\kk_{2}\kk_{3}} 
w_{\kk_{1},\kk_{2},\kk_{3}} \delta_{\kk_{1}+\kk_{2}+\kk_{3}=0}
\left( {\cal A}_{\kk_{1}}{\cal A}_{\kk_{2}}{\cal A}_{\kk_{3}}\right.\\
& & \left.
-3\sigma_{z}g_{\kk_{1}}{\cal A}_{\kk_{2}}{\cal A}_{\kk_{3}}
+3g_{\kk_{1}}g_{\kk_{2}}{\cal A}_{\kk_{3}}
-\sigma_{z}g_{\kk_{1}}g_{\kk_{2}}g_{\kk_{3}}  \right),
\end{eqnarray*}
where ${\cal A}_{\kk}=\bek+\bemkd$.
The first term is the anharmonic coupling between the new modes,
the third describes a shift of the oscillator
equilibria, the
fourth is a shift of the energy levels that may be included in
the energy $E$. 

Of interest here is the second term that describes 
two-phonon absorption,
emission and scattering. Only the latter may lead to
energy-conserving processes that do not involve real transitions
between system states.
Using commutation relations and symmetries
of the anharmonic coefficients one may write the relevant
(scattering) part of the anharmonic Hamiltonian
\begin{eqnarray}\label{h1s}
H_{1}^{(\mathrm{s})}&=&
-\sigma_{z}\sum_{\kk,\qq}w_{\qq-\kk,-\qq,\kk}g_{\bm{q}-\bm{k}}\\
\nonumber
&&\times
\left[ 
\beqd\bek-\delta_{\qq,\kk}n_{\kk}
+\delta_{\qq,\kk}\left(n_{\kk}+1/2\right) \right].
\end{eqnarray}
The last term may again be included into the energy $E$ (it
vanishes in the limit of infinite reservoir volume). It should be
noticed that Eq.~(\ref{h1s}) is non-diagonal in phonon modes which
allows for real phonon scattering processes (unlike the model of
Ref. \cite{goupalov01}). A similar scattering Hamiltonian may be
obtained by including higher exciton levels into a purely harmonic
model \cite{muljarov04}.

It is known 
that scattering on a heavy Brownian particle leads
to decoherence and localization of the quantum state of the latter
\cite{joos85,gallis90,hornberger03}. In order to see that the same is
true in the present case of scattering of bosonic modes on a two-state
quantum system and to extract the corresponding long-time behavior
we write the evolution equation with the interaction Hamiltonian
(\ref{h1s}). Assuming that the time scales of the dephasing
process are longer than those related to the
reservoir memory and initial dephasing one may
consistently describe the long time dynamics in the Markov
limit. From the resulting Lindblad equation one finds the solution
describing exponential pure dephasing at long times with the rate
\cite{epaps}
\begin{eqnarray}\label{T2}
\frac{1}{T_{2}} & = & 2\int_{-\infty}^{\infty}dt
\langle B^{\dag}(t)B(0)\rangle \\\nonumber
& = & 4\pi \sum_{\kk,\qq}|w_{\qq-\kk,-\qq,\kk}|^{2}|g_{\qq-\kk}|^{2}
n_{\kk}(n_{\qq}+1)\delta(\wq-\wk),
\end{eqnarray}
where $B(t)=\sum_{\kk,\qq}w_{\qq-\kk,-\qq,\kk}g_{\qq-\kk}
[ \beqd\bek e^{i(\wq-\wk)t} -\delta_{\qq,\kk}n_{\kk} ]$
are reservoir operators with vanishing equilibrium average.
The dephasing rate is finite at $T>0$, which 
should be contrasted with the harmonic case, where the dephasing
effect vanishes in the Markov limit. Thus, 
reservoir anharmonicity has qualitatively changed the
decoherence properties of the system.

The rate given by Eq.~(\ref{T2}) is consistent with the
scattering picture described in the introduction. A boson 
scatters from the state $\kk$ to $\qq$. The scattering amplitude 
is proportional to the magnitude of the displacement field, governed
by the system-reservoir coupling constants $g_{\qq-\kk}$, 
and to the anharmonic coupling $w_{\qq-\kk,-\qq,\kk}$. The momentum
transfer in such an event cannot exceed the inverse size of the
displacement field which will be reflected by a cutoff in 
$g_{\qq-\kk}$ (see below). The scattering probability depends on the 
occupations of the initial and final states. Finally,
since no real transitions between system states are allowed, the
scattering must be elastic, as expressed by the energy conserving
Dirac delta.

In order to see if this effect may be of importance under realistic
conditions, let us now study the 
specific case of a single electron in
a pair of semiconductor quantum dots, as in the recent coherent
manipulation experiment \cite{hayashi03}. 
The decay of coherence between the localized states corresponds in
this case to a localization process, hence $T_{2}$ may be referred to
as the \textit{localization time}.
Since we are interested in the dephasing due to anharmonicity, we
disregard phonon-assisted tunneling between the
states ($\sigma_{x,y}$ coupling) which might appear
only if
the states overlap (such terms obviously lead to exponential
decoherence; the anharmonicity effects for such real transition
processes were studied elsewhere \cite{verzelen00,jacak02a-jacak03a}).
The Hamiltonian is therefore
\begin{eqnarray*}
H_{\mathrm{QD}}&=&\sum_{i=0,1}|i\rl i|\left[ 
\epsilon_{i}+\sumk\fk^{(i)*}(\tilde{b}_{\kk}+\tilde{b}^{\dag}_{-\kk}) 
\right] \\
&&+\sumk\hbar\wk\tilde{b}_{\kk}^{\dag}\tilde{b}_{\kk},
\end{eqnarray*}
where $|0\rangle,|1\rangle$ are the basis states (each localized 
in one of the two dots),
$\epsilon_{i}$ are the energies of the two states,
$\tilde{b}_{\kk}$ are phonon operators (with respect to the
unperturbed equilibrium), and $\fk^{(i)}$ are coupling constants. 
This Hamiltonian is transformed to the form of Eq.~(\ref{h0}) by the
canonical transformation (shift) of the phonon modes, 
$\tilde{b}_{\kk}=\bk+(\fk^{(0)}+\fk^{(1)})/(2\hbar\wk)$. The effective
system-reservoir coupling constants are then
$\Fk=\fk^{(0)}-\fk^{(1)}$. This shift modifies also the anharmonic
Hamiltonian $H_{1}$ but, since the transformation is
independent of system state, no extra coupling will appear. The
new linear and quadratic terms in $H_{1}$ may
be removed by re-diagonalizing the phonon Hamiltonian, which produces
negligible higher order corrections to the couplings.

In polar semiconductors, the strongest lattice displacement
(\textit{polaron} \cite{mahan00}) is related to
longitudinal optical (LO) phonons, which are also subject to strong
anharmonic coupling to acoustic phonons \cite{vallee91,barman02}. 
Therefore, the present discussion will be restricted to the scattering
of longitudinal acoustic (LA) phonons on the LO displacement field. 
We assume dispersionless
LO modes with frequency $\Omega=54$ ps$^{-1}$
(all values for GaAs \cite{adachi85}).
The electron wave functions will be
modelled by isotropic Gaussians of size $L$ and the
QDs will be displaced by a distance $D$ along $z$ \footnote{ 
The full discussion, including all coupling channels and more
realistic wave functions will be
presented elsewhere.}. 
The physical coupling constants
are \cite{jacak02a-jacak03a} 
\begin{equation}\label{fk}
\fk^{(0,1)}=\frac{e}{\hbar k}
\sqrt{\frac{\hbar\Omega}{2V\varepsilon_{0}\tilde{\varepsilon}}}
e^{-\left( \frac{Lk}{2} \right)^{2}}e^{\pm i\frac{k_{z}D}{2}}.
\end{equation}
Hence
\begin{eqnarray}\label{gk}
|\gk|^{2} & = & \frac{|\fk^{(0)}-\fk^{(1)}|^{2}}{(\hbar\Omega)^{2}}\\
\nonumber
& = &
\frac{2e^{2}}{k^{2}\hbar^{3}V\varepsilon_{0}\tilde{\varepsilon}\Omega}
e^{-\frac{1}{2} (Lk)^{2}}\sin^{2}\frac{k_{z}D}{2},
\end{eqnarray}
where $e$ is the electron charge, $V$
the normalization volume of the phonon modes, $\varepsilon_{0}$
the vacuum dielectric constant, and $\tilde{\varepsilon}=70$ the
lattice part of the relative dielectric constant. The
Gaussian momentum cutoff reflects the momentum conservation and
momentum-position uncertainty for an electron wave packet of size $L$.

In an anharmonic process, an
LO phonon interacts with two LA phonons with
linear dispersion $\wk=ck$ up to the Debye wave vector
$k_{\mathrm{D}}=11$ nm$^{-1}$, where $c=5150$ m/s is the speed of sound.
For this process, the general form of the coupling is 
$w_{\qq-\kk,\qq,\kk}=(w_{0}/\sqrt{V})\sqrt{qk}$ \cite{LL10}, where
$\qq,\kk$ pertain to the LA phonons and we neglected the dependence
on the LO phonon momentum in the narrow range of
its relevant values $k\lesssim 1/L\ll k_{\mathrm{D}}$. 
Using the measured lifetime $\tau_{0}=9.2$ ps of the LO
phonon at $\kk=0$ \cite{vallee91}
one finds, using the Fermi golden rule
with the anharmonic Hamiltonian (\ref{h1}),
$w_{0}^{2}=\frac{64\pi\hbar^{2}c^{5}}{\tau_{0}\Omega^{4}}$.

Substituting the above result along with Eq.~(\ref{gk}) 
into Eq.~(\ref{T2}) one finds after some algebra 
\cite{epaps}
\begin{eqnarray}\label{T2-expli}
\frac{1}{T_{2}} & = & \frac{64}{\pi^{2}}
\frac{e^{2}(k_{\mathrm{B}}T)^{5}}{\tau_{0}\hbar^{6}\Omega^{5}
\varepsilon_{0}\tilde{\varepsilon}c}
\int_{0}^{\infty}\frac{dx}{x}e^{-x^{2}}
\left(1-\frac{\sin\alpha x}{\alpha x}\right) \\
\nonumber
&&\times \left[ 
\phi(x_{\mathrm{D}}) 
- \phi\left( \frac{xx_{\mathrm{D}}}{\sqrt{2}k_{\mathrm{D}}L} \right) \right],
\end{eqnarray}
where $x_{\mathrm{D}}=(\hbar c k_{\mathrm{D}})/(k_{\mathrm{B}}T)$, 
$\alpha=\sqrt{2}D/L$, and 
\begin{displaymath}
\phi(x)=\int_{0}^{x}duu^{5}\frac{e^{u}}{(e^{u}-1)^{2}}.
\end{displaymath}

\begin{figure}[tb] 
\begin{center} 
\unitlength 1mm
\begin{picture}(80,30)(0,6)
\put(0,0){\resizebox{80mm}{!}{\includegraphics{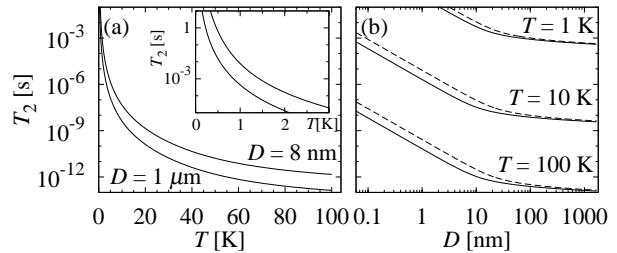}}}
\end{picture} 
\end{center} 
\caption{\label{fig:decoh}
Left: the dependence of the localization time on temperature for $L=4$
nm (inset shows the low-temperature sector). Right: the
dependence on the distance between the states for $L=4$ nm (solid
lines) and $L=8$ nm (dashed).}
\end{figure}

The dependence of the dephasing time $T_{2}$ on temperature and
distance between the basis states is shown in
Fig.~\ref{fig:decoh}. Note that at moderate temperatures 
$T_{2}$ is much longer than the
initial dephasing and reservoir memory times ($\sim 1$ ps) so that the
results are consistent.
For low temperatures the localization rate
is limited by the number of occupied initial states and the resulting 
allowed final states (due to energy conservation). 
As a result, one finds $T_{2}\sim
T^{-7}$ and the effect is extremely weak for sub-Kelvin
temperatures (see Fig.~\ref{fig:decoh}a). 
However, already at $T\sim 20$ K the dephasing rate is
of the order of nanoseconds and becomes comparable to typical
coherent manipulation times on such structures. At $\sim 100$
K the dephasing time drops to several picoseconds even for closely
spaced dots. This temperature dependence is much stronger
than in the quantum Brownian motion 
\cite{caldeira85,zurek03}. This is not astonishing,
since this feature depends on the reservoir density of states
and the physical nature of the coupling
so that no universality can be expected here. In fact, similar strong
temperature dependence has been found for localization
processes due to scattering of light on dielectric balls and on free 
electrons \cite{joos85}.

At all temperatures the
localization rate is increased by over an order of magnitude when the
distance $D$ between the dots grows from nanometers to micrometers. 
This distance dependence is shown in detail in
Fig.~\ref{fig:decoh}b. It turns out that the dephasing rate grows
rather fast ($\sim D^{2}$) as long as the wave functions overlap. 
This dependence is a general feature in the regime of ``ineffective
single scattering event'' \cite{joos85}. When
the states get separated, the increase of the localization rate
continues, although it becomes only logarithmic. This is
related to the crossover to the regime of spatially distinct states where a
single scattering event is sufficient to extract the position information.
The dependence on the dot size $L$ becomes less and less
important as the distance grows, as shown by the comparison between
the dot sizes of 4 and 8 nm. This should be contrasted with the 
harmonic model \cite{krummheuer05}, where the dependence on the
separation saturates while that on the size does not. 
Therefore, the present results cannot be
fully explained by merely invoking the known effect of increased
carrier-phonon coupling for separated carriers.
The 
logarithmic asymptotic behavior is unexpected on the grounds of
a general discussion \cite{gallis90,hornberger03}; here it results
from the long range nature of carrier-phonon interaction, manifested
by the long wavelength singularity in Eq.~(\ref{fk}).

The presented results show that anharmonicity of a super-Ohmic bosonic
reservoir leads to a qualitative change in the dynamics of an open
system.
Exponential (Markovian) pure dephasing appears even though only
partial decoherence was present without anharmonicity. For typical
coupling properties, the dephasing rate depends very strongly on
temperature. In the case of a
superposition between two spatially separated states the rate of the resulting
localization grows also with the spatial distance between the two
states.

It should be stressed that the dephasing (localization) mechanism
described here is inherent to physical properties of the system
and appears universally for any localized states embedded in a
translationally invariant bosonic reservoir with a certain degree of
anharmonicity. As such, it sets material-dependent limits to 
system coherence, independent on any design improvements that might
eliminate noise sources that dominate dephasing in the current
experiments \cite{hayashi03}.

The present result is of importance to a few areas. First,
it describes an additional dephasing mechanism that must be taken into
account both in design of devices relying on quantum coherence and
in interpretation of experiments. Apart from the localization
effect discussed here, the anharmonic scattering mechanism will
contribute, e.g., to the broadening of the zero-phonon line in QD
spectroscopy \cite{borri01,bayer02}.
Second, the dependence of
dephasing on the distance in space may affect scalability of
quantum computing schemes and applicability of concatenation
techniques used in quantum fault-tolerant architectures \cite{knill05}. 
Third, appearance of dephasing in models with non-singular, super-Ohmic
coupling may be of importance for emerging of classicality from 
quantum evolution \cite{zurek03}.

The author is grateful to M. Axt, T. Kuhn, L. Jacak, and A. Janutka for
discussions and to the Alexander von Humboldt Foundation and Polish
MNI (PBZ-MIN-008/P03/2003, PB 2 P03B 08525) for support.  


\end{document}


\author{Pawe{\l} Machnikowski}
\affiliation{Institute of Physics, Wroc{\l}aw University of
Technology, 50-370 Wroc{\l}aw, Poland}

\title{Derivation of the $T_{2}$ time}

\begin{abstract}
This auxiliary material explains how the long time scale behavior of
the system was extracted in the Markov limit and gives the technical
details of derivation of the final formula [Eq.~(8)] 
for the dephasing times in the QD localization example.
\end{abstract}

\maketitle

The system evolution in the Markovian limit is found following the
usual approach \cite{breuer02}.
The Hamiltonian $H_{1}^{(\mathrm{s})}$ (discarding the last term)
is written in the interaction picture in the form 
\begin{displaymath}
H_{1}^{(\mathrm{s})}(t)= -\sigma_{z}\otimes B(t),
\end{displaymath}
where $B(t)$ is the reservoir operator defined below Eq.~(5).
We assume that the initial state of the system is separable in terms
of the shifted operators $\beta_{\kk}$,
\begin{displaymath}
\rho(0)=\rho_{\mathrm{S}}(0)\otimes\rho_{\mathrm{B}},
\end{displaymath}
where $\rho_{\mathrm{B}}$ is the thermal equilibrium state of the reservoir.
Physically, this corresponds
to preparing the system in such a way that the polaronic displacement field is
formed by adiabatically following the system evolution. This
assumption corresponds to typical timescales in the system of two QDs
used as an example in the paper but, in any case, the long-time behavior
cannot depend on the way the initial system state was prepared.

The evolution equation for the density matrix of the total system
is written in the usual integral form and the reduced density matrix of
the two-level system is then obtained by tracing out the reservoir degrees
of freedom,
\begin{displaymath}
\dot{\rho}_{\mathrm{S}}(t)=-\int_{0}^{t}d\tau \tr_{\mathrm{B}}
\left[ H_{1}^{(\mathrm{s})},
\left[ H_{1}^{(\mathrm{s})},\rho(\tau) \right]  \right],
\end{displaymath}
where we used the fact that the operator $B(t)$ has a vanishing
average at thermal equilibrium so that the 1st order term vanishes.

At this point, a series of standard approximations is introduced, leading to
the Born-Markov equation for the evolution of the two-level
system. First, one assumes that the perturbation of the reservoir
state due to the joint evolution of the two subsystems is weak.
It may then be assumed
that at each instant of time the system state is separable, with
the reservoir at thermal equilibrium,
$\rho(s)=\rho_{\mathrm{S}}(s)\otimes\rho_{\mathrm{B}}$,
which leads to the Nakajima-Zwanzig equation
\begin{displaymath}
\dot{\rho}_{\mathrm{S}}(t)=-\int_{0}^{t}d\tau \tr_{\mathrm{B}}
\left[ H_{1}^{(\mathrm{s})},
\left[
H_{1}^{(\mathrm{s})},\rho_{\mathrm{S}}(\tau)\otimes\rho_{\mathrm{B}} 
\right]  \right]
=2\int_{0}^{t}d\tau \re\langle B(t-\tau)B(t)\rangle 
\left(\sigma_{z}\rho_{\mathrm{S}}(\tau)\sigma_{z}-\rho_{\mathrm{S}}(\tau)
\right), 
\end{displaymath}
where we use the fact that, for a hermitian operator $B(t)$, one has 
$\langle B(t)B(t')\rangle=\langle B(t')B(t)\rangle^{*}$.
Next, one assumes that the evolution of the two-level system is very
slow and its state changes negligibly during the reservoir memory time 
$\tau_{\mathrm{M}}$ over which
the correlation function $\langle B(\tau)B\rangle$ is essentially nonzero.
This allows us to replace $\rho_{\mathrm{S}}(\tau)$ by
$\rho_{\mathrm{S}}(t)$. Finally, at times $t\gg\tau_{\mathrm{M}}$ the
lower integration limit may be set to $-\infty$. Using the fact that
$\rho_{\mathrm{B}}$ is a stationary state, the Born-Markov evolution equation
for the reduced density matrix of the two-level system may be written
as
\begin{displaymath}
\dot{\rho}_{\mathrm{S}}(t)=\frac{1}{2}\gamma
\left[\sigma_{z}\rho_{\mathrm{S}}(t)\sigma_{z}-\rho_{\mathrm{S}}(t)
\right],
\end{displaymath}
where 
\begin{displaymath}
\gamma=4\re\int_{0}^{\infty}\langle B(t)B\rangle dt
=2\int_{-\infty}^{\infty}\langle B(t)B\rangle dt.
\end{displaymath}
It is easy to see that the diagonal elements of $\rho_{\mathrm{S}}(t)$
remain constant, while the non-diagonal ones decay with the time
constant $T_{2}=1/\gamma$. The final formula of Eq.~(5) follows in a
straightforward manner by explicitly calculating the correlation
function using the definition of the operator $B(t)$, as given below
Eq.~(5) in the paper.

This result may be directly used to calculate the $T_{2}$ time constant for
the localization process used as an example in the paper. Inserting
the coupling constants $g_{\qq-\kk}$ and $w_{\qq-\kk,-\qq,\kk}$ into
Eq.~(5) and changing both summations into integrals according to the
usual formula (for normalization volume $V$),
$\sumk\to V\int d^{3}k/(2\pi)^{3}$, one gets
\begin{displaymath}
\gamma=\frac{1}{T_{2}}=\frac{8e^{2}c^{5}}{
\pi^{4}\tau_{0}\Omega^{5}\hbar\varepsilon_{0}\tilde{\varepsilon}}
\int d^{3}k\int d^{3}q \frac{kq}{(\kk-\qq)^{2}}
e^{-\frac{L^{2}}{2}(\kk-\qq)^{2}}\sin^{2}\frac{(k_{z}-q_{z})D}{2}
\delta(ck-cq)n_{\kk}(n_{\qq}+1).
\end{displaymath}
Let us introduce the new variable $\pp=\kk-\qq$. Next, we express
$\qq$ in spherical coordinates $(q,\theta,\phi)$ 
in the reference system with the $z'$
axis along $\pp$ and $\pp$ itself in spherical coordinates 
$(p,\vartheta,\varphi)$  in the
original (absolute) reference
frame. In the new variables the Dirac delta is transformed into 
\begin{displaymath}
\delta(ck-cq)=\frac{1}{c}\delta(q-|\qq+\pp|)
=\frac{\delta(p+2q\sin\theta)}{c|\sin\theta|}.
\end{displaymath}
The result becomes
\begin{eqnarray*}
\gamma & = & \frac{32e^{2}c^{5}}{
\pi^{2}\tau_{0}\Omega^{5}\hbar\varepsilon_{0}\tilde{\varepsilon}}
\int_{q<k_{\mathrm{D}}} dqq^{2}\int_{-\pi/2}^{\pi/2}d\theta \cos\theta
\int_{|\qq+\pp|<k_{\mathrm{D}}} dpp^{2} 
	\int_{-\pi/2}^{\pi/2}d\vartheta \cos\vartheta\\
&&\times
\frac{q|\pp+\qq|}{p^{2}}
e^{-\frac{1}{2}(Lp)^{2}}\sin^{2}\frac{pD\sin\vartheta}{2}
\frac{\delta(p+2q\sin\theta)}{c|\sin\theta|}
n_{\qq+\pp}(n_{\qq}+1),
\end{eqnarray*}
where the integration with respect to $\phi,\varphi$ has been performed.
It is now possible to integrate over $p$ and $\vartheta$. 
Introducing the new variables 
\begin{displaymath}
x=\frac{\hbar c q}{k_{\mathrm{B}}T},\;\;\;\;
t=\frac{\sqrt{2}k_{\mathrm{B}}TLx}{\hbar c}\sin\theta
\end{displaymath}
we can write the result in the form
\begin{displaymath}
\gamma=\frac{64e^{2}c^{4}}{
\pi^{2}\tau_{0}\Omega^{5}\hbar\varepsilon_{0}\tilde{\varepsilon}}
\left( \frac{k_{\mathrm{B}}T}{\hbar c} \right)^{5}
\int_{0}^{x_{\mathrm{D}}} dx x^{5}\frac{e^{x}}{(e^{x}-1)^{2}}
\int_{0}^{\sqrt{2}k_{\mathrm{D}}Lx/x_{\mathrm{D}}}\frac{dt}{t}
e^{-t^{2}}\left( 1-\frac{\sin\alpha t}{\alpha t} \right),
\end{displaymath}
where $x_{\mathrm{D}}=\hbar c k_{\mathrm{D}}/(k_{\mathrm{B}}T)$ and 
$\alpha=\sqrt{2}D/L$.
This formula may be integrated by parts to yield
\begin{displaymath}
\gamma=\frac{64e^{2}c^{4}}{
\pi^{2}\tau_{0}\Omega^{5}\hbar\varepsilon_{0}\tilde{\varepsilon}}
\left( \frac{k_{\mathrm{B}}T}{\hbar c} \right)^{5}
\left[  
\phi(x_{\mathrm{D}})
\int_{0}^{\sqrt{2}k_{\mathrm{D}}L}\frac{dx}{x}
e^{-x^{2}}\left( 1-\frac{\sin\alpha x}{\alpha x} \right)
-\int_{0}^{\sqrt{2}k_{\mathrm{D}}L}\frac{dx}{x}
e^{-x^{2}}\left( 1-\frac{\sin\alpha x}{\alpha x} \right)
\phi\left( \frac{xx_{\mathrm{D}}}{\sqrt{2}k_{\mathrm{D}}L} \right) 
\right].
\end{displaymath}
Since $k_{\mathrm{D}}L\gg 1$ the upper limits of both integrals may be
extended to $\infty$, which leads to the form of Eq.~(8).

\bibliographystyle{prstyea}
\bibliography{abbr,quantum}